\documentclass[a4paper,10pt,twoside]{cpc-hepnp}

\usepackage{multicol}
\usepackage{graphicx}
\usepackage{booktabs}
\usepackage{amssymb,bm,mathrsfs,bbm,amscd}
\usepackage[tbtags]{amsmath}
\usepackage{lastpage}
\usepackage{CJK}

\begin{document}
\begin{CJK*}{GB}{gbsn}
\CJKtilde



\title{Role of magnetic fields on the outer crust in a magnetar}

\author{Jiang Wei(½¯Íþ)$^1$ and Chen Yan-jun(³Âê̾ü)$^{1,2;1)}$,\email{chenyjy@ustc.edu.cn} } \maketitle

\address{
$^1$ Department of Physics and Electronic Science, Changsha
University of Science and Technology, Changsha, Hunan 410114,
China\\ $^2$ Hunan Provincial Key Laboratory of Flexible Electronic
Materials Genome Engineering, Changsha, Hunan 410114, China }

\begin{abstract}
We explore the properties of 4110 nuclides from $Z=5$ to $Z=82$ with
the Sky3D code and the composition of the outer crust in the
magnetars under extreme magnetic fields. The effects of the
variation of the nuclear masses due to the magnetic fields on the
outer crust are comprehensively studied. The neutron-drip transition
pressure, the equation of state and neutron fraction in the outer
crust have also been discussed.
\end{abstract}




\begin{multicols}{2}

\section{Introduction}

The origins of soft $\gamma$-ray repeaters and some anomalous x-ray
pulsars are generally associated with a kind of special neutron
stars, the so-called magnetars, which possess very strong magnetic
fields (MF) of the order of 10$^{14}$$-$10$^{15}$ G on the surface
\cite{dun92,vie07,mer08,tie13,ola14,kas17}. The accurate description
of the crust of a magnetar is essential to understand some
astrophysical phenomena involving magnetars, such as the
oscillations \cite{car06,wat07,sam07,and09,pas14,nan16} and cooling
\cite{agu09,vig13,pot18} in magnetars. In the present work, we will
focus on the study on the outer crust in magnetars, where the fully
ionized atoms are thought to be arranged in a Coulomb lattice
embedded in a gas of degenerate electrons \cite{cha08} and the high
MF can have non-negligible influence on the composition and equation
of state due to the Landau-Rabi quantization of electron motion
\cite{lai91,cha12,cha15,mut19,par23}. Moreover, the MF with
$B\gtrsim10^{17}$ G may modify appreciably the nuclear structure due
to the alteration of the size and even the ordering of energy levels
\cite{kon00,kon01,pen11,ste16}. The calculations about the outer
crust in neutron stars are based on the method proposed by Baym,
Pethick, and Sutherland \cite{baym71}, where the masses of nuclei
act as input parameters. Until now, there are very few studies about
the effects of the change of nuclear masses in the presence of MF on
the outer crust \cite{bas15}. It is mainly due to the fact that the
surface fields of magnetars inferred from the current observations
are up to a few times 10$^{15}$ G \cite{mer08,kas17} and $3.2 \times
10^{16}$ G from the latest work \cite{sob23}, and the fields with
such order of magnitude play a small role in the nuclear masses.
Nonetheless, considering the upper limit of the interior magnetic
fields being the order of 10$^{18}$ G according to the virial
theorem \cite{lai91} and numerical simulations
\cite{car01,kiu082,fri12,pili,chatt15,tso22}, it is still worth
investigating the nuclear masses under the extreme high magnetic
fields and the corresponding effects on the crustal properties in
magnetars prior to possible astronomical observations.

For studying the properties of nuclei in the presence of magnetic
fields, we use the publicly available Sky3D code based on the Skyrme
density functional theory \cite{mar14,sch18} in the present work to
solve the time-independent Hartree-Fock (HF) equations on a
three-dimensional grid without further symmetry assumptions. This
code has been used to study a wide range of problems, such as the
nuclear structure, collective vibrational excitations and heavy-ion
collisions.

This article is organized as follows. In Sec. 2, we describe the 
formulas for the present work. In Sec. 3, the calculated results and
some discussions are given. Finally, the summary is present in Sec.
4.

\section{ The Formalism}

The energy eigenvalue of an electron under MF presents the
Landau-Rabi levels whose energies are given by
    \begin{equation}\label{eq4}
        E_{\nu} = \sqrt{c^2p_z^2 + m_e^2c^4(1 + 2\nu B_{\star})}\,\,,
    \end{equation}
where $c$ is the speed of light, $m_e$ is the rest mass of the
electron, $p_z$ is the component of the momentum along the field,
$\nu=n+1/2+s/2=0,1,2,...$ with $n$ being the principal quantum
number and $s$ the spin along the magnetic field axis with $+1$ for
spin-up and $-1$ for spin-down cases. Note that $\nu=0$ corresponds
to one single spin state while all $\nu>0$ states correspond to two
spin states. Here $B_{\star}=B/B_c$ is the magnetic field $B$
measured in units of the critical field $B_c$ defined as
    \begin{equation}\label{eq3}
        {B_c} = \frac{m_e^2c^3}{e\hbar} \approx 4.41 \times {10^{13}}\,\rm{G}\,\,,
    \end{equation}
at which the electronic cyclotron energy reaches the electron rest
mass energy.

In the case of temperature $T=0$ K, the Fermi momenta of electrons
$k_{Fe}$ for different numbers of $\nu$ are related to the
electronic chemical potential $\mu_e$ in the magnetar's crust in
equilibrium as $\mu_e^2= c^2p_{Fe}(\nu)^2 + m_e^2c^4(1 + 2\nu
B_{\star})$. The requirement that $p_{Fe}\geq 0$ determines the
maximum number of $\nu$ labeled as $\nu_{max}$. Then the expression
of the electron density under MF can be given by
\begin{equation}\label{eq9}
n_e=\frac{{{B_\star}}}{{2{\pi
^2}\lambda_e^3}}\sum\limits_{\nu=0}^{{\nu_{\max }}}
{{g_{\nu}}{x_e}(\nu)}\,\,,
    \end{equation}
where $\lambda_e=\hbar/{m_e}c$ is the Compton wavelength, the
degeneracy $g_{\nu}$ is 1 for $\nu=0$ and 2 for $\nu\geq 1$ and the
dimensionless Fermi momentum is defined by
$x_e(\nu)=p_{Fe}(\nu)/m_ec$. The electronic energy density and
pressure are respectively given by
    \begin{eqnarray}
       && \varepsilon_e = \frac{{{B_\star}{m_e}}}{{2{\pi ^2}\lambda _e^3}}\sum\limits_{\nu = 0}^{{\nu_{\max }}} {{g_{\nu}}(1 + 2\nu{B_\star}){\psi _ + }\left[ {\frac{{{x_e}(\nu)}}{{\sqrt {(1 + 2\nu{B_\star})} }}}
        \right]}\,\,,\nonumber\\ \label{eq10}\\
       &&          P_e = \frac{{{B_\star}{m_e}}}{{2{\pi ^2}\lambda _e^3}}\sum\limits_{\nu = 0}^{{\nu_{\max }}} {{g_{\nu}}(1 + 2\nu{B_\star}){\psi _ - }\left[ {\frac{{{x_e}(\nu)}}{{\sqrt {(1 + 2\nu{B_\star})} }}} \right]}\,\,,\nonumber\\\label{eq11}
\end{eqnarray}
    where ${\psi _ \pm }(x) = x\sqrt {1 + {x^2}}  \pm \ln (x + \sqrt {1 + {x^2}}
    )$. Also the results in the absence of the MF ($B = 0$ G) corresponding to Eqs. (\ref{eq10}) and (\ref{eq11}) are given in this work by
\begin{eqnarray}\label{eq14}
&&\varepsilon_e = \frac{{{m_e}}}{{8{\pi ^2}\lambda _e^3}} \left\{
x{(1 + {x^2})^{\frac{1}{2}}}(1 + {x^2}) - \ln [x + {(1 +
{x^2})^{\frac{1}{2}}}]\right\}\,\,,\nonumber\\ \\
&&{P_e} = \frac{{{m_e}}}{{8{\pi ^2}\lambda _e^3}}\left\{ x{(1 +
{x^2})^{\frac{1}{2}}}(\frac{2}{3}x^2 - 1) + \ln [x +
{(1 + {x^2})^{\frac{1}{2}}}]\right\}\,\,,\nonumber\\
    \end{eqnarray}
where $x=\lambda_e(3\pi^2n_e)^{1/3}$.

For $B=0$ G case, the ions are believed to be arranged in a
body-centered cubic lattice, the lattice energy per baryon is
approximately given by \cite{bas15}
    \begin{equation}\label{eq16}
        \varepsilon_l= - 3.40665 \times {10^{ - 3}}\frac{Z^2}{A^{4/3}}p_{F_b}\,\,,
    \end{equation}
where the baryonic Fermi momentum $p_{F_b}$ is in MeV. The
associated lattice pressure is $P_l= \varepsilon_l/3$. The total
pressure is therefore $P=P_e+P_l$. Following the arguments in Ref.
\cite{bas15}, this work still use Eq. (\ref{eq16}) to give the
contribution of the lattice energy in the presence of magnetic
fields.

The Gibbs free energy per nucleon can be written as \cite{bas15}
\begin{equation}\label{eq21}
g=\frac{M(A,Z,B)}{A} +
\frac{Z}{A}{\mu_e}+\frac{{4Z}}{{An_e}}{P_l}\,\,,
\end{equation}
where $M(A,Z,B)$ is the mass of the nuclei with proton number $Z$
and atomic number $A$, and is the function of $B$ because we will
consider the effects of the MF on nuclear masses in this work. Then
the distribution of nuclei in the outer crust of neutron stars in
equilibrium can be obtained by searching a series of nuclei for the
minimum of these Gibbs free energies.

In this work, we follow the line of Ref. \cite{ste16} where the
Sky3D code is used to calculate the masses of nuclei under the
MF. 
The effects of external MF are included by introducing the
B-field-related hamilton operators $\hat{H}_{p,n}^{(B)}$ into the
original Hamiltonian in the Sky3D code, which are given by
\begin{eqnarray}
&&\hat{H}_p^{(B)} = - \frac{e}{2m_pc}(\vec L + g_p\vec S) \cdot \vec B\,\,,\;\;\;\;\;\textrm{for proton},\label{eq19}\\
&&\hat{H}_n^{(B)} = - \frac{e}{2m_nc}{g_n}\vec S \cdot \vec
B\,\,,\;\;\;\;\;\textrm{for neutron},\label{eq20}
\end{eqnarray}
where $m_p$ and $m_n$ are the proton and neutron masses,
respectively, $g_p=5.5856$ and $g_n=-3.8263$ are the g-factors of a
proton and neutron, and $\vec L$ and $\vec S$ are the orbital and
spin angular momenta, respectively.

The following are some settings for the Sky3D code: we consider, in
an isolated system, 24 grid points in each direction in
three-dimensional Cartesian coordinates, with a distance of $1.0\
\text {fm}$ between each grid point. The force chosen in the type of
Skyrme force is 'SLy6',  which is mainly applied to neutron-rich
nuclei and neutron matter in the field of astrophysics
\cite{ch1998}. For the setting of the Newtonian gradient iteration
method, the damping step is adjusted to $x_0=0.4$ and the damping
adjustment parameter to $E_0 = 100$. The initial radii of the
harmonic oscillator state are set to $3.1\ \text {fm}$. We begin the
calculation in the absence of MF, which is used as the input values
to obtain the results at $B=10^{16}$ G. Then these results at
$B=10^{16}$ G are used for the higher MF and similar procedures are
carried until $B=10^{18}$ G. The maximum number of iterations are
set to 1500 and the convergence criterion is that the energy
fluctuation is less than $10^{-6}\ \text {MeV}$.

\section{ Results and Discussion}

In Figs.~\ref{figg1}-~\ref{figg4}, the calculations on 4110 nuclides
from $Z=5$ to $Z=82$, ranging from neutron fraction 0.54 to 0.73,
are made for the MF with $B=0$ G (\textit{left panels}) and
$B=10^{18}$ G (\textit{right panels}). Since the strong MF break the
pairing in nuclei, these figures and the remainder of this work
neglect the pairing energy in $B\neq 0$ G cases for simplicity.
Comparing Fig.~\ref{figg1}(a) and (b), we see that this super-strong
MF increase the binding energies $\textit{BE}$s of these nuclei by
8\%$\sim$15\%. At $B=10^{18}$ G, shown in Fig.~\ref{figg2}, the root
mean square radii $r_{\text {rms}}$ are seen to be larger than in
the absence of MF.

The MF also play an important role in the shape of the nucleus. In
this work, we show 
the total deformation $\beta$ and the triaxiality $\gamma$ in
Figs.~\ref{figg3}-~\ref{figg4}, respectively. Here the parameters
$\beta$ and $\gamma$ are known as Bohr-Mottelson parameters
\cite{gr1996}. $\beta = 0$ refers to a spherical nucleus, which is
non-deformed. When $\beta$ is non-zero, the nucleus is deformed. The
shape of a nucleus is prolate if $\gamma=0 ^\circ$, oblate if
$\gamma=60 ^\circ$, and between the prolate and oblate deformations
for $0 ^\circ<\gamma<60 ^\circ$. In Fig. ~\ref{figg3}(a), the
nuclides in the region near the magic numbers have $\beta$ = 0,
which means they are non-deformed. But in Fig. ~\ref{figg3}(b), this
phenomenon does not appear and almost all the nuclides are deformed.
The low $\beta$ regions ($\beta <0.05$) mainly focus on the areas
with $N$=26$\sim$36, 52$\sim$68 and 96$\sim$112. Note that some
regions with large $\beta$ in the absence of MF may become the
small-$\beta$ ones in the strong MF, e.g., the blue area around
$Z=58-68$ and $N=92-102$ in Fig. ~\ref{figg3}(a), where $\beta$ is
greater than 0.2, can not be seen in Fig. ~\ref{figg3}(b). Thus, the
strong MF may cause some nuclei with large deformation to become
more 'spherical'. Compared with the results in Fig.~\ref{figg4}(a),
the triaxialities vary considerably in Fig.~\ref{figg4}(b), and
interestingly, we see $\gamma=0^\circ$ around $N=44, 74, 114, 172$,
between which $\gamma=60^\circ$ shows up. This means that in the
super-strong MF, in general, the prolate and oblate deformations of
a nucleus emerge alternately with increasing neutron numbers.

In Fig.~\ref{figg5}, we calculate the binding energy \textit{BE},
radius $R$, and deformation parameters $\beta$ and $\gamma$ of six
nuclides which may exist in the outer crust of a cold nonaccreting
neutron star as a function of magnetic field strength. What is the
most obvious in this figure is that $\textit{BE}$s are not simply
positively correlated with the strength of the magnetic field, but
rise jaggedly. It is because the spin-up and spin-down states split
under the MF, and the energy levels of a nucleus may undergo the
rearrangement for B $\gtrsim 10^{17}$ G~\cite{ste16}. The analogue
in atomic physics is the transition from Zeeman to Paschen-Back
effect when the MF become very strong. In this figure, we see that
$R$, $\beta$ and $\gamma$ are nearly unchanged for $B\lesssim
4.1\times 10^{17}$ G, but $\textit{BE}$s still increase. At
$B\gtrsim 4.1\times 10^{17}$ G, $R$s increase with increasing MF,
while $\beta$s rise initially then have small changes above some
certain values of magnetic field strength. In addition, at $B\gtrsim
4.1\times 10^{17}$ G, the phenomenon that $\gamma$s become
$\gamma=0^\circ$ rapidly when the MF increase, then increase to
about $\gamma=30^\circ$ and subsequently return to $\gamma=0^\circ$
at higher MF implies that the nuclei may not become more prolate
with higher MF and might relate to the alternate emergence of the
prolate and oblate deformation of a nucleus with increasing neutron
numbers, seen in Fig.~\ref{figg4}(b).


To understand the jaggedness of the $\textit{BE}$ in
Fig.~\ref{figg5} more thoroughly, we display in Fig.~\ref{figg6} the
neutron and proton energy levels of ${}^{208}$Pb, for $B=0$ G,
$B=10^{17}$ G and $B=10^{18}$ G. The interaction between the MF and
the magnetic moment for $B\neq 0$ splits the energy levels. The
neutrons have no orbital magnetic moment, thus the gaps between the
neutron magic numbers are still obvious at $B=10^{17}$ G, and higher
MF ($B=10^{18}$ G here) are needed to modify or even eliminate the
neutron shell structure. On the other hand, the effects of the MF on
the protons are enough due to the orbital magnetic moment so that
some outer and inner energy levels in the absence of MF cross over
at $B=10^{17}$ G and more energy levels cross over at $B=10^{18}$ G.
This rearrangement of energy levels leads to the appearance of the
jaggedness for the $\textit{BE}$.

\begin{center}
\includegraphics[width=9cm]{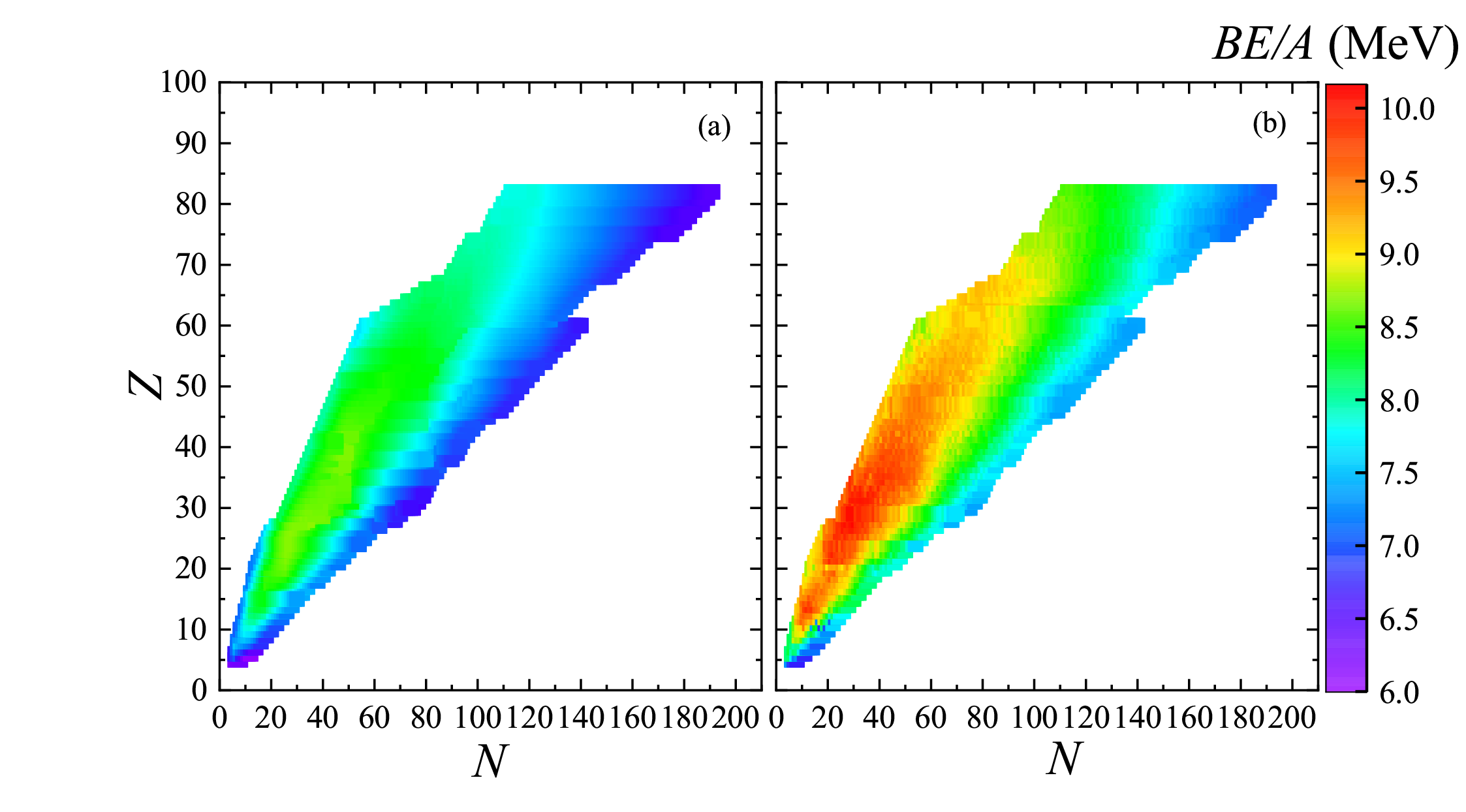}
\figcaption{\label{figg1}  Binding energy per nucleon
$\textit{BE}/A$ (in MeV) for the nuclides with the neutron ($N$) and
proton ($Z$) numbers at (a) $B = 0$ G (\textit{left panel}) and (b)
$B= 10^{18}$ G (\textit{right panel}).}
\end{center}

\begin{center}
\includegraphics[width=9cm]{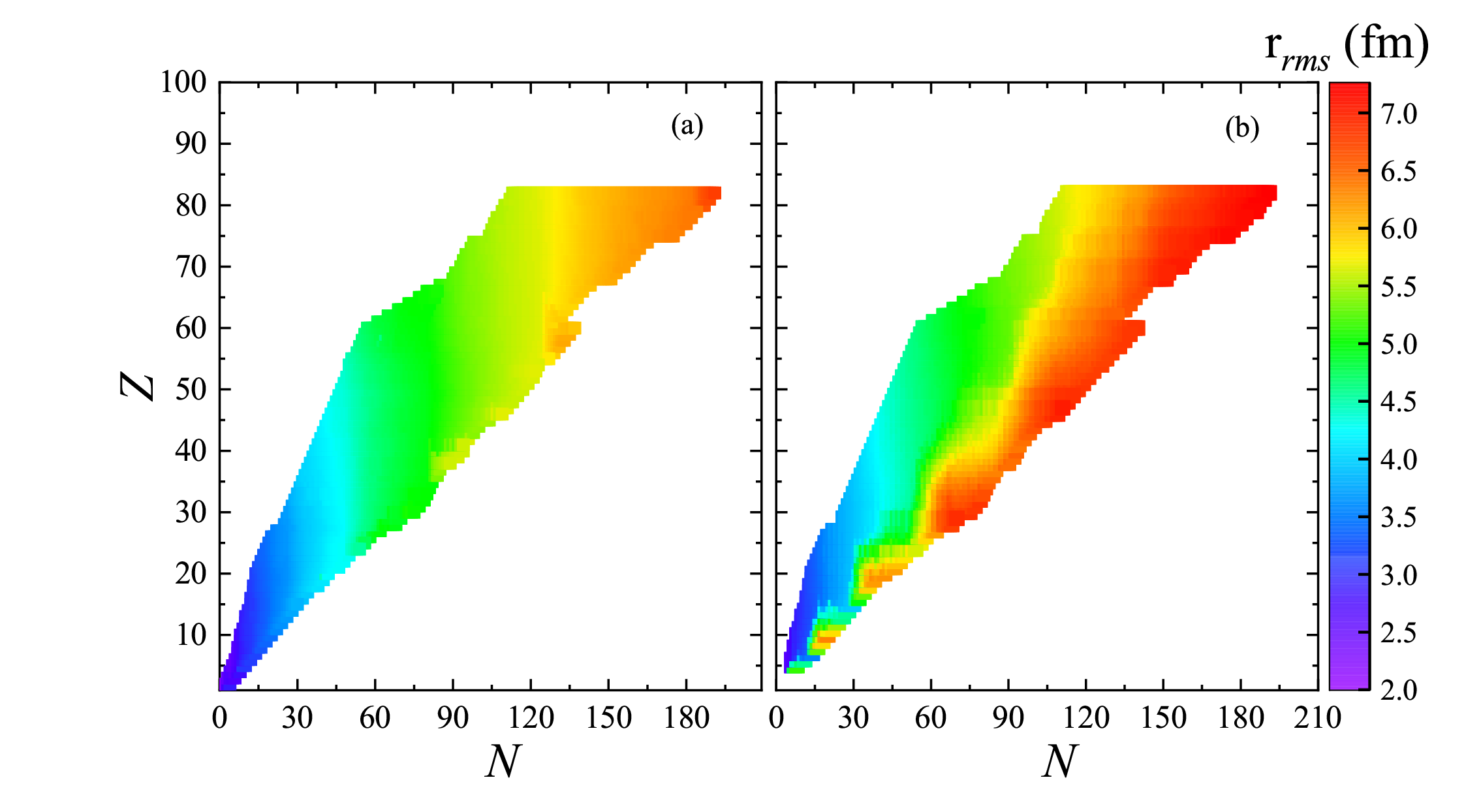}
\figcaption{\label{figg2}  Root mean square radii $r_{\text {rms}}$
(in fm) for the nuclides with the neutron ($N$) and proton ($Z$)
numbers at (a) $B = 0$ G (\textit{left panel}) and (b) $B= 10^{18}$
G (\textit{right panel}). }
\end{center}

\begin{center}
    \includegraphics[width=9cm]{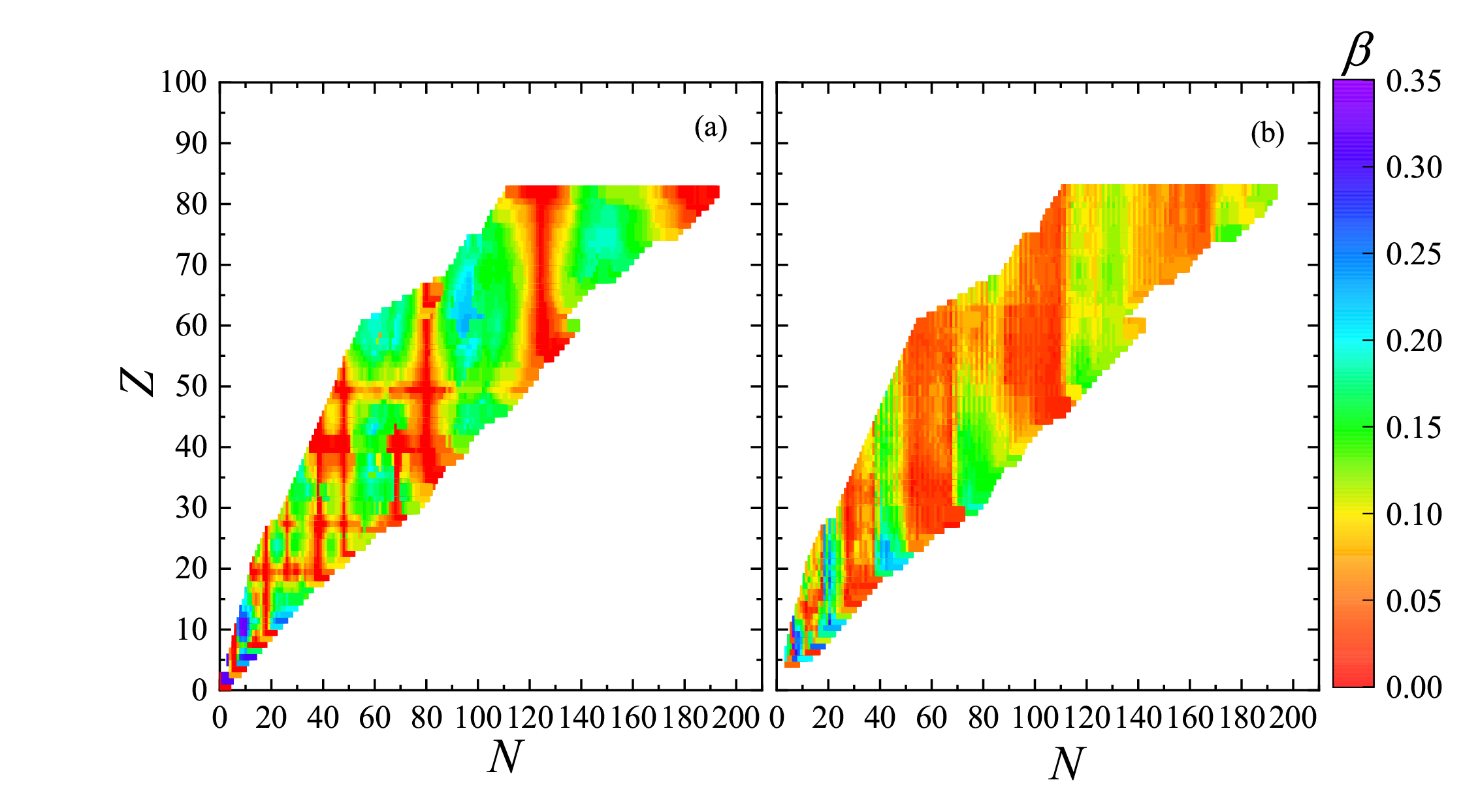}
    \figcaption{\label{figg3} Total deformation $\beta$ for the nuclides with the neutron ($N$) and proton ($Z$)
numbers at (a) $B = 0$ G (\textit{left panel}) and (b) $B= 10^{18}$
G (\textit{right panel}).}
\end{center}

\begin{center}
    \includegraphics[width=9cm]{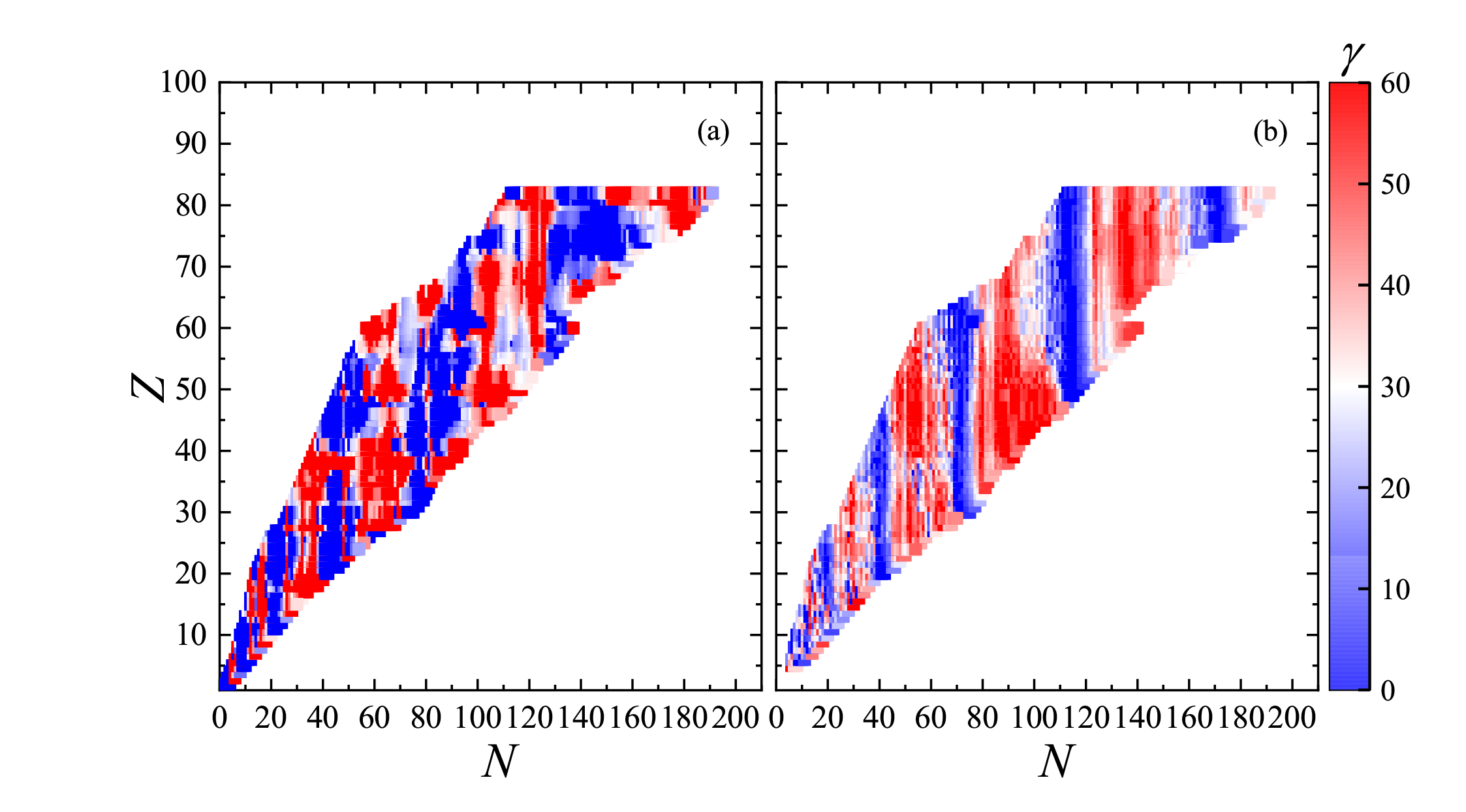}
    \figcaption{\label{figg4}  Triaxiality $\gamma$ for the nuclides with the neutron ($N$) and proton ($Z$)
numbers at (a) $B = 0$ G (\textit{left panel}) and (b) $B= 10^{18}$
G (\textit{right panel}). }
\end{center}

\begin{center}
\includegraphics[width=8cm]{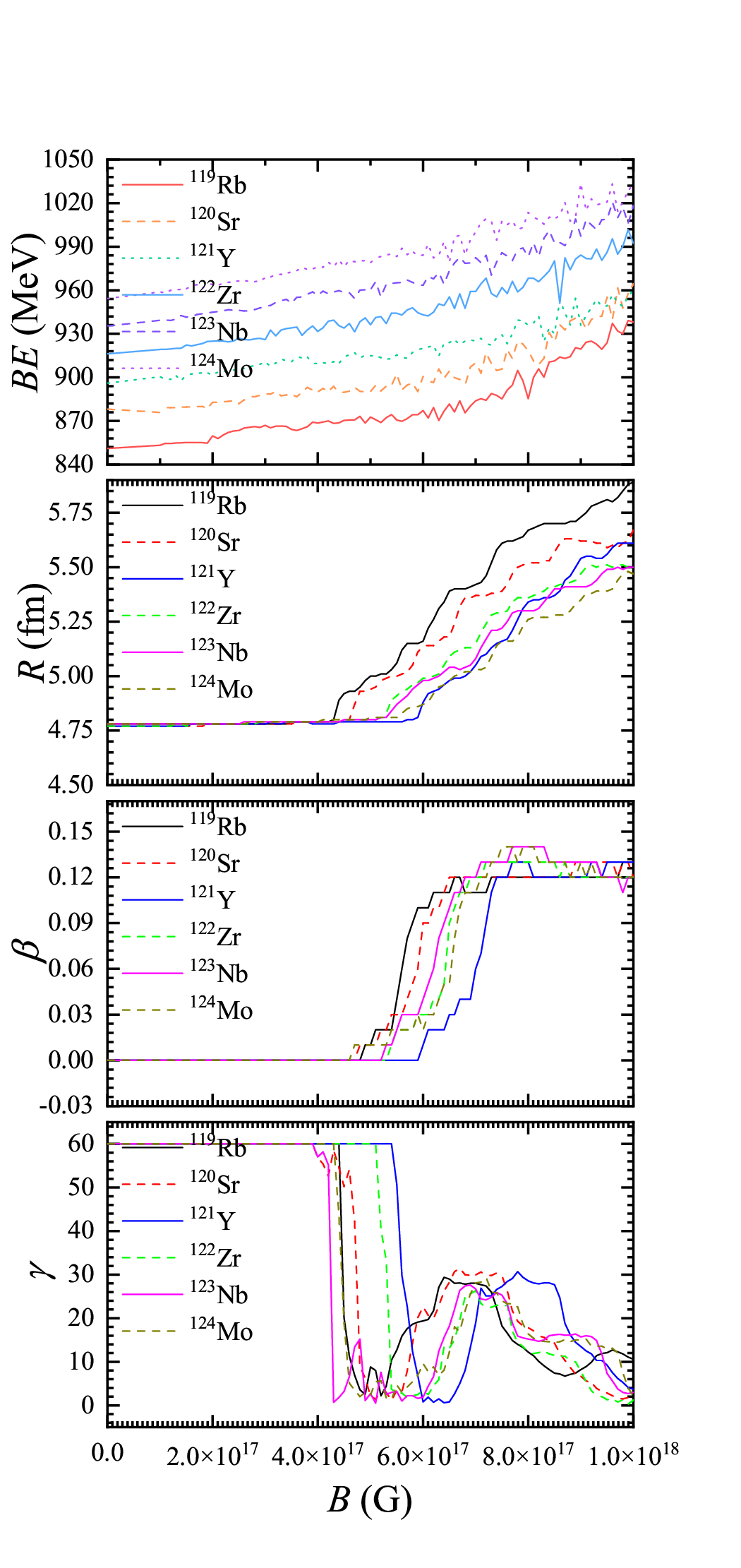}
\figcaption{\label{figg5} Binding energy (\textit{BE}), radius $R$,
and deformation parameters $\beta$ and $\gamma$ of six nuclides
which may exist in the outer crust of a cold nonaccreting neutron
star as a function of magnetic field strength. }
\end{center}

\begin{center}
    \includegraphics[width=8cm]{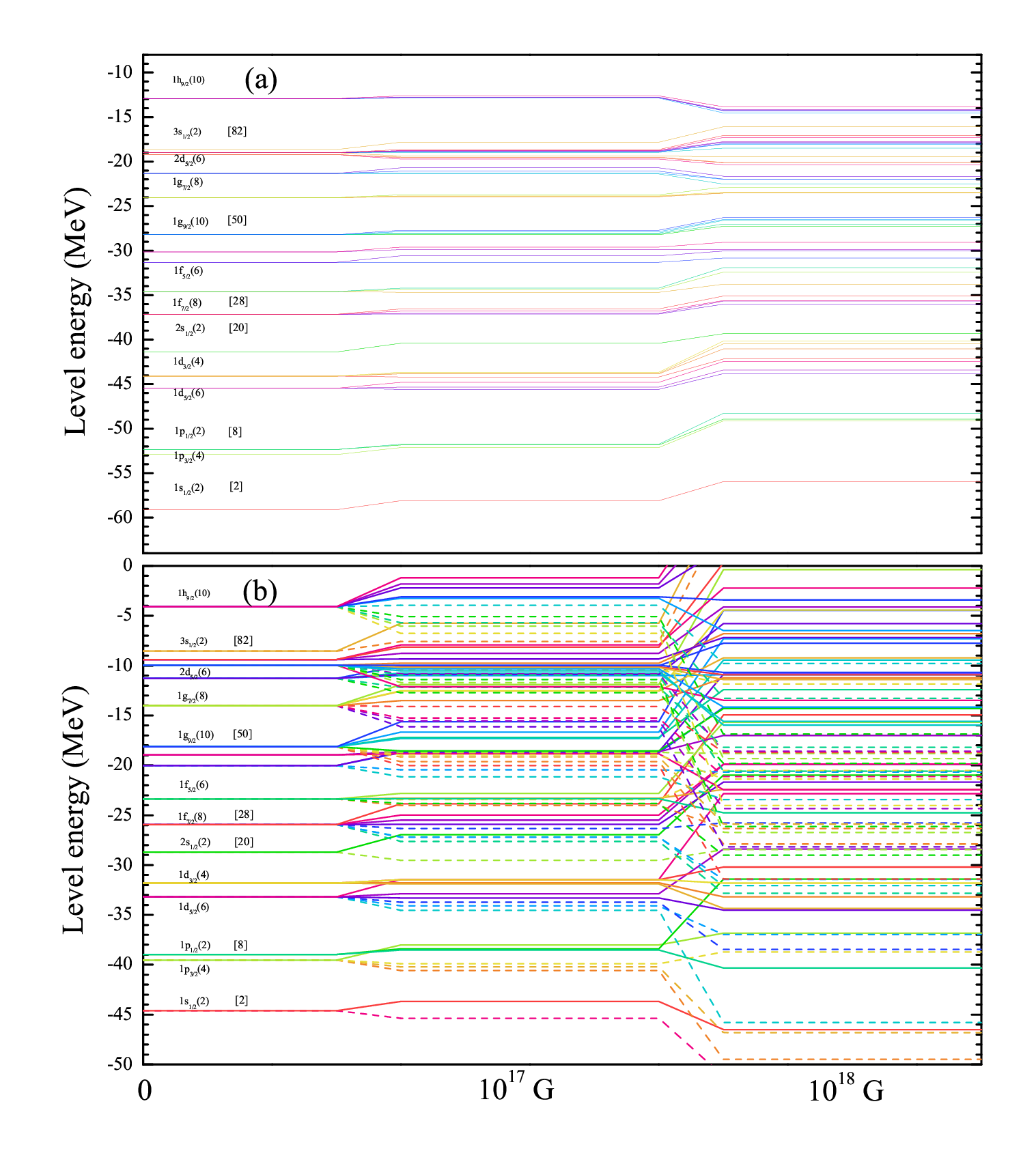}
    \figcaption{\label{figg6} Energy levels of (a) neutron (\textit{upper panel}) and (b) proton (\textit{lower panel}) (in MeV) of ${}^{208}$Pb at $B=0$ G, $B=10^{17}$ G and $B=10^{18}$ G.}
\end{center}

\begin{center}
\includegraphics[width=9cm]{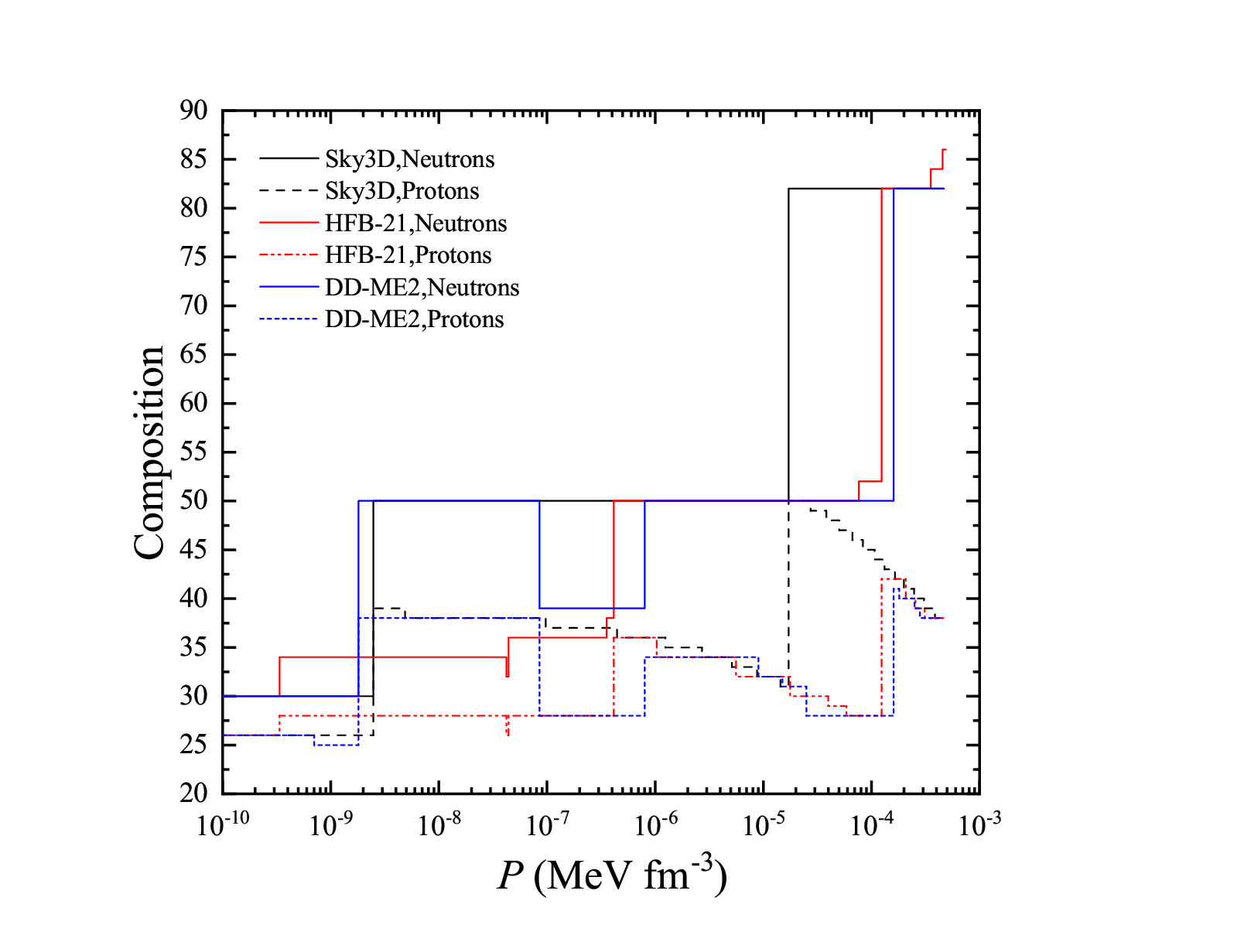}
\figcaption{\label{figg7}  Composition of the outer crust of a cold
nonaccreting neutron star in the absence of a magnetic field $(B=0\
\text G)$ where the nuclear masses are obtained with the Skyrme
force SLy6 \cite{ch1998} by Sky3D code, the Hartree-Fock-Bogoliubov
method labeled by HFB-21 \cite{cha12} and the relativistic
mean-field model labeled by DD-ME2 \cite{bas15}. }
\end{center}

\begin{center}
\includegraphics[width=9cm]{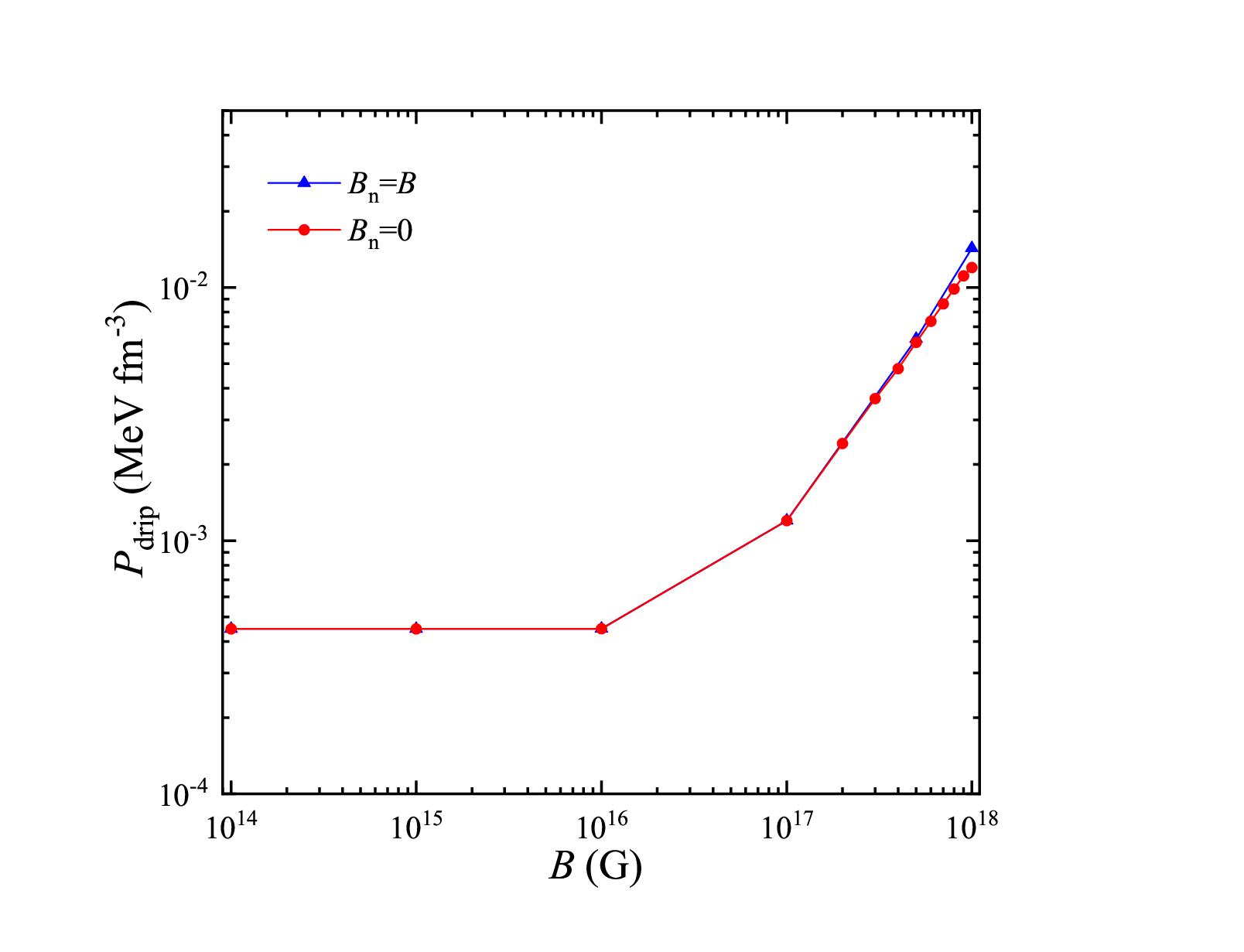}
\figcaption{\label{figg8} Neutron-drip transition pressure
$P_{\text{drip}}$ versus magnetic field strength $B$, where the
effect of the magnetic fields on the nuclear masses is included
($B_{\text n}=B$) and not included $(B_{\text n}=0)$. }
\end{center}

In Fig.~\ref{figg7}, we illustrate the equilibrium composition of
the outer crust in the absence of MF by minimizing Eq. (\ref{eq21}).
The nuclear masses are calculated with the Skyrme force SLy6
\cite{ch1998} by Sky3D code. For comparison, the results based on
the Hartree-Fock-Bogoliubov (HFB) method labeled by HFB-21 and the
relativistic mean-field model labeled by DD-ME2 are also shown in
this figure, respectively \cite{cha12,bas15}. We see some trends in
common for these three models. The equilibrium element in the
outermost part of crust is $^{56}$Fe. With the increasing pressure,
two constant plateaus of neutron number appear, which correspond to
two magic numbers, i.e., $N=50$ and $82$, respectively. Accompanied
by these two plateaus is the obvious lift of proton number with the
subsequent decrease. All the maximum pressures for these models are
similar $\sim 4.7\times 10^{-4}$ MeV fm$^{-3}$, at which the
neutrons begin to drip out of the nuclei, indicating the boundary
between the outer and inner crusts of the neutron stars.
Nonetheless, there are some model-dependent features for these
models, e.g., DD-ME2 displays a dip for the neutron and proton
numbers in the interval between $P\sim 10^{-7}-10^{-6}$ MeV
fm$^{-3}$ and the $N=50$ plateau begins to appear in HFB-21 at
higher pressure.

Fig.~\ref{figg8} displays the neutron-drip transition pressure
$P_{\text{drip}}$ as a function of the magnetic field, where
$P_{\text{drip}}$ is obtained by requiring Gibbs energy in
Eq.~(\ref{eq21}) equaling to the neutron mass. It is shown that the
effects of MF on $P_{\text{drip}}$ are almost nothing other than for
$B>10^{16}$ G. Furthermore, we see that when $B\gtrsim10^{17}$ G,
$P_{\text{drip}}$ is nearly a linear function of magnetic field.
These results are in agreement with that in Refs~\cite{cha12,bas15}.
In addition, the variation on the nuclear masses due to the MF
($B_n=B$) plays a ignorable role for $B\lesssim10^{17}$ G because
the MF with such orders of magnitude can not affect the nuclei
obviously. For the extreme high MF, $B=10^{18}$ G, $P_{\text{drip}}$
in the $B_n=B$ case increases about 20\% in comparison with that in
the case without taking the change of the nuclear masses into
account ($B_n=0$).

\begin{center}
\includegraphics[width=8cm]{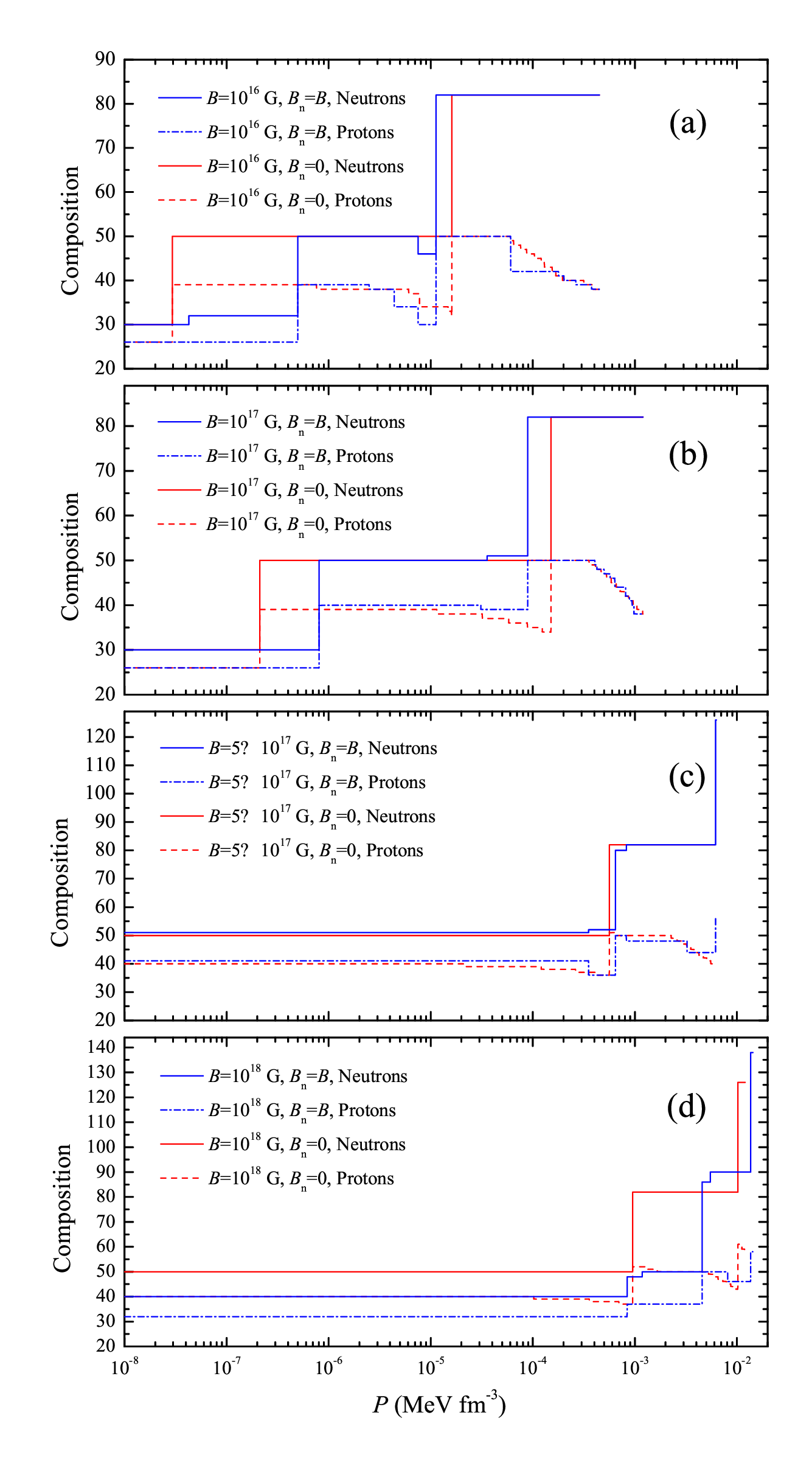}
\figcaption{\label{figg9} Composition of the outer crust of a cold
nonaccreting magnetar for the four cases, (a) $B = {10^{16}}$ G, (b)
$B = {10^{17}}$ G, (c) $B=5\times 10^{17}$ G and (d) $B = {10^{18}}$
G, where the effect of the magnetic fields on the nuclear masses is
included ($B_{\text n}=B$) and not included $(B_{\text n}=0)$. }
\end{center}

In Fig.~\ref{figg9}, we illustrate the equilibrium composition of
the outer crust for (a) $B= 10^{16}$ G, (b) $B= 10^{17}$ G, (c)
$B=5\times 10^{17}$ G, and (d) $B= 10^{18}$ G. The variation on the
nuclear masses due to the MF is considered ($B_{\text n} = B$) and
not considered ($B_{\text n} = 0$). The nuclear masses are affected
slightly by the MF for $B\leq 10^{17}$ G, but we can still find
obvious discrepancy between the $B_{\text n} = B$ and $B_{\text n} =
B$ cases in Fig.~\ref{figg9} (a) and Fig.~\ref{figg9} (b). However,
this result may be model dependent \cite{bas15}. It can be seen that
the plateaus with $N=50$ and $82$ are present in Fig.~\ref{figg9}
(a) and (b). There is a clear shift for the $N=50$ plateau to higher
pressures in comparison with that in the absence of MF seen in
Fig.~\ref{figg1}, but the pressure at which the $N=82$ plateau
begins to emerge does not change very much for $B=10^{16}$ G. This
is in agreement with the results in Ref.~\cite{bas15}. We also show
the composition in detail for $B= 10^{16}$ G and $B= 10^{17}$ G in
Table.~\ref{table2} and \ref{table3}, where only the $B_{\text n} =
B$ case is listed. In both tables, from $P_{\text{ion}}$ to the
maximum pressure in the outer crust $P_{\text{drip}}$, $^{56}$Fe
occurs first, then $^{89}$Y on the $N=50$ plateau and $^{132}$Sn,
$^{124}$Mo, $^{123}$Nb, $^{122}$Zr, $^{120}$Sr on the $N=82$ plateau
are present. We notice that the same nuclides emerge at higher
pressures in the stronger MF. It is mainly due to the reduced
electronic Fermi energy under high MF. In addition, some nuclides
may appear or disappear under the different magnetic field strength.
$^{88}$Sr and $^{84}$Se existing in the $B=10^{16}$ G case disappear
on the $N=50$ plateau for $B=10^{17}$ G, while $^{90}$Zr appear on
this plateau. The nuclides not seen on the $N=82$ plateau for
$B=10^{16}$ G, i.e., $^{131}$In, $^{130}$Cd, $^{129}$Ag, $^{128}$Pd
and $^{126}$Ru can exist in the $B=10^{17}$ G case. Moreover, we
find that the nuclide at the neutron-drip transition in
Fig.~\ref{figg9} (a) and (b) is same, i.e., $^{120}$Sr, which is
also one in the absence of MF (not shown in this paper). It is
consistent with the observation in Ref.~\cite{cha12}, but it is
$^{124}$Sr there.

For $B= 5\times10^{17}$ G and $B= 10^{18}$ G, the composition in the
neutron star outer crust is significantly changed. Firstly we focus
on the $B_{\text n} = 0$ case. Fig.~\ref{figg9} (c) and (d) show
that the outermost nuclide in the outer crust is not $^{56}$Fe, but
$^{90}$Zr, which is just on the $N=50$ platform, and spans in the
broad range of pressures. The plateau with $N=82$ is still present
and delays to higher pressures compared with that in the weak MF. It
is surprising that the neutron plateau with new magic number $N=126$
can be seen for $B= 10^{18}$ G. This is in disagreement with the
conclusion in Ref.~\cite{bas15}, where the authors preset a
medium-heavy nuclide as the upper limit of the composition in the
outer crust. It is also pointed out by Ref.~\cite{sek23} where the
HFB method is used and the effect of the MF on the nuclear masses is
not included. Therefore, for the super-strong MF, such neutron-rich
nuclide might be present in the outer crust. In the $B_{\text n} =
B$ case, we see from Fig.~\ref{figg9} (c) and Table \ref{table4}
that for $B=5\times10^{17}$ G the outermost nuclide is not on the
$N=50$ platform, i.e., $^{92}$Nb, and the nuclide at
$P_{\text{drip}}$ is $^{182}$Ba, which is just on the $N=126$
plateau. For stronger MF, i.e., $B=10^{18}$ G ($B_{\text n} = B$),
Fig.~\ref{figg4} (d) and Table \ref{table5} show that the platforms
with $N=82$ and $N=126$ disappear. The outermost nuclide is
$^{72}$Ge, and the nuclide at $P_{\text{drip}}$ is $^{196}$Ce which
differs evidently from $N=126$ plateau. It results from the fact
that the binding energies rise jaggedly with increasing MF due to
the rearrangement of the energy levels for $B> 10^{17}$ G, as shown
in Fig.~\ref{figg5}. Thus, some concepts, e.g., magic numbers,
should be reexamined when the effects of MF on nuclear masses are
taken into account.

\begin{center}
\tabcaption{\label{table2}Composition of the outer crust of a
magnetar at $B=10^{16}$ G. $P_{\text {min}}$ ($P_{\text{max}}$) is
the minimum (maximum) pressure of the appearing nuclide, in units of
$ \text{MeV fm}^{-3}$, at which $n_{\text {min}}$ ($n_{\text
{max}}$), in units of $ \text{fm}^{-3}$, is the corresponding
average minimum (maximum) baryon number density. $P_{\text {ion}}$
and $n_{\text {ion}}$ are the complete ionization pressure and
density. Here only the $B_{\text n}=B$ case which includes the
effect of the magnetic fields on the nuclear masses is
shown.}\footnotesize
\begin{tabular*}{85mm}{lcccc}
        \hline
        Nucleus           & $P_{\text {min}}$           & $P_{\text {max}}$           & $n_{\text {min}}$   & $n_{\text {max}}$   \\ \hline
        ${}_{26}^{56}$Fe  & $P_{\text{ion}}$    & $4.27\cdot 10^{-8}$ & $n_{\text{ion}}$    & $5.54\cdot 10^{-7}$ \\
        ${}_{26}^{58}$Fe   & $4.28\cdot 10^{-8}$ & $4.98\cdot 10^{-7}$ & $5.74\cdot 10^{-7}$ & $1.53\cdot 10^{-6}$ \\
        ${}_{39}^{89}$Y   & $4.99\cdot 10^{-7}$ & $2.50\cdot 10^{-6}$ & $1.59\cdot 10^{-6}$ & $3.34\cdot 10^{-6}$ \\
        ${}_{38}^{88}$Sr  & $2.51\cdot 10^{-6}$ & $4.40\cdot 10^{-6}$ & $3.39\cdot 10^{-6}$ & $4.44\cdot 10^{-6}$ \\
        ${}_{34}^{84}$Se  & $4.41\cdot 10^{-6}$ & $7.51\cdot 10^{-6}$ & $4.62\cdot 10^{-6}$ & $5.99\cdot 10^{-6}$ \\
        ${}_{30}^{76}$Zn  & $7.52\cdot 10^{-6}$ & $1.13\cdot 10^{-5}$ & $6.28\cdot 10^{-6}$ & $7.63\cdot 10^{-6}$ \\
        ${}_{50}^{132}$Sn  & $1.14\cdot 10^{-5}$ & $6.07\cdot 10^{-5}$ & $8.03\cdot 10^{-6}$ & $3.64\cdot 10^{-5}$ \\
        ${}_{42}^{124}$Mo  & $6.08\cdot 10^{-5}$ & $1.79\cdot 10^{-4}$ & $3.80\cdot 10^{-5}$ & $1.07\cdot 10^{-4}$ \\
        ${}_{41}^{123}$Nb  & $1.80\cdot 10^{-4}$ & $2.02\cdot 10^{-4}$ & $1.09\cdot 10^{-4}$ & $1.18\cdot 10^{-4}$ \\
        ${}_{40}^{122}$Zr  & $2.03\cdot 10^{-4}$ & $2.64\cdot 10^{-4}$ & $1.20\cdot 10^{-4}$ & $1.49\cdot 10^{-4}$ \\
        ${}_{39}^{121}$Y  & $2.65\cdot 10^{-4}$ & $3.77\cdot 10^{-4}$ & $1.51\cdot 10^{-4}$ & $1.91\cdot 10^{-4}$ \\
        ${}_{38}^{120}$Sr  & $3.79\cdot 10^{-4}$ & $4.49\cdot 10^{-4}$ & $1.95\cdot 10^{-4}$ & $2.31\cdot 10^{-4}$ \\
        \hline\end{tabular*}
\end{center}

\begin{center}
\tabcaption{\label{table3}Same as Table \ref{table2}, but at
$B=10^{17}$ G.}\footnotesize
\begin{tabular*}{85mm}{lcccc}
        \hline
        Nucleus           & $P_{\text {min}}$           & $P_{\text {max}}$           & $n_{\text {min}}$   & $n_{\text {max}}$   \\ \hline
       ${}_{26}^{56}$Fe  & $P_{\text{ion}}$    & $8.07\cdot 10^{-7}$ & $n_{\text{ion}}$    & $7.54\cdot 10^{-6}$ \\
       ${}_{40}^{90}$Zr  & $8.08\cdot 10^{-7}$ & $3.10\cdot 10^{-5}$ & $8.26\cdot 10^{-6}$ & $3.75\cdot 10^{-5}$ \\
       ${}_{39}^{89}$Y  & $3.11\cdot 10^{-5}$ & $3.58\cdot 10^{-5}$ & $3.81\cdot 10^{-5}$ & $4.06\cdot 10^{-5}$ \\
       ${}_{39}^{90}$Y  & $3.59\cdot 10^{-5}$ & $8.99\cdot 10^{-5}$ & $4.12\cdot 10^{-5}$ & $6.38\cdot 10^{-5}$  \\
       ${}_{50}^{132}$Sn  & $9.00\cdot 10^{-5}$ & $4.05\cdot 10^{-4}$ & $7.35\cdot 10^{-5}$ & $1.52\cdot 10^{-4}$ \\
       ${}_{49}^{131}$In  & $4.06\cdot 10^{-4}$ & $4.24\cdot 10^{-4}$ & $1.54\cdot 10^{-4}$ & $1.58\cdot 10^{-4}$ \\
       ${}_{48}^{130}$Cd  & $4.25\cdot 10^{-4}$ & $5.01\cdot 10^{-4}$ & $1.60\cdot 10^{-4}$ & $1.74\cdot 10^{-4}$ \\
       ${}_{47}^{129}$Ag  & $5.02\cdot 10^{-4}$ & $5.71\cdot 10^{-4}$ & $1.76\cdot 10^{-4}$ & $1.87\cdot 10^{-4}$ \\
       ${}_{46}^{128}$Pd & $5.72\cdot 10^{-4}$ & $6.38\cdot 10^{-4}$ & $1.90\cdot 10^{-4}$ & $2.00\cdot 10^{-4}$ \\
       ${}_{44}^{126}$Ru & $6.39\cdot 10^{-4}$ & $8.13\cdot 10^{-4}$ & $2.06\cdot 10^{-4}$ & $2.33\cdot 10^{-4}$ \\
       ${}_{42}^{124}$Mo & $8.14\cdot 10^{-4}$ & $8.86\cdot 10^{-4}$ & $2.40\cdot 10^{-4}$ & $2.50\cdot 10^{-4}$ \\
       ${}_{41}^{123}$Nb & $8.87\cdot 10^{-4}$ & $9.41\cdot 10^{-4}$ & $2.54\cdot 10^{-4}$ & $2.62\cdot 10^{-4}$ \\
       ${}_{40}^{122}$Zr & $9.42\cdot 10^{-4}$ & $9.79\cdot 10^{-4}$ & $2.66\cdot 10^{-4}$ & $2.71\cdot 10^{-4}$ \\
       ${}_{38}^{120}$Sr & $9.80\cdot 10^{-4}$ & $1.20\cdot 10^{-3}$ & $2.81\cdot 10^{-4}$ & $3.10\cdot 10^{-4}$ \\
        \hline\end{tabular*}
\end{center}

In Fig. \ref{figg10}, we compare the behavior of the pressure $P$ as
a function of the baryonic density $n$ for the three magnetic field
cases. The result for $B=0$ G is also shown for comparison. It can
be seen that the effects of the variation on the nuclear masses due
to the MF are negligible. In the low density region, the phenomenon
that the density is almost unchanged over a wide range of pressures
is more obvious under higher MF. It indicates that the outermost
material of the magnetar is almost incompressible. This is because
the electrons occupy only the lowest Landau level at low densities.
On the other hand, the electrons may occupy a growing number of
levels rapidly with increasing densities, and the effects of MF will
become unimportant. It may result in the similar slope of these
lines at higher densities to that for $B=0$ G which is approximately
linear, as can be seen in this figure.

\begin{center}
\tabcaption{\label{table4}Same as Table \ref{table2}, but at
$B=5\times 10^{17}$ G.}\footnotesize
\begin{tabular*}{85mm}{lcccc}
        \hline
        Nucleus           & $P_{\text {min}}$           & $P_{\text {max}}$           & $n_{\text {min}}$   & $n_{\text {max}}$   \\ \hline
        ${}_{41}^{92}$Nb  & $P_{\text{ion}}$    & $3.53\cdot 10^{-4}$ & $n_{\text{ion}}$    & $2.85\cdot 10^{-4}$ \\
        ${}_{36}^{88}$Kr  & $3.54\cdot 10^{-4}$ & $6.44\cdot 10^{-4}$ & $3.09\cdot 10^{-4}$ & $4.10\cdot 10^{-4}$ \\
        ${}_{50}^{130}$Sn  & $6.45\cdot 10^{-4}$ & $8.27\cdot 10^{-4}$ & $4.42\cdot 10^{-4}$ & $4.97\cdot 10^{-4}$ \\
        ${}_{48}^{130}$Cd  & $8.28\cdot 10^{-4}$ & $3.25\cdot 10^{-3}$ & $5.17\cdot 10^{-4}$ & $1.00\cdot 10^{-3}$ \\
        ${}_{44}^{126}$Ru  & $3.26\cdot 10^{-3}$ & $6.16\cdot 10^{-3}$ & $1.05\cdot 10^{-3}$ & $1.44\cdot 10^{-3}$ \\
        ${}_{56}^{182}$Ba  & $6.17\cdot 10^{-3}$ & $6.27\cdot 10^{-3}$ & $1.65\cdot 10^{-3}$ & $1.66\cdot 10^{-3}$ \\
        \hline\end{tabular*}
\end{center}

\begin{center}
\tabcaption{\label{table5}Same as Table \ref{table2}, but at
$B=10^{18}$ G.}\footnotesize
\begin{tabular*}{85mm}{lcccc}
        \hline
        Nucleus           & $P_{\text {min}}$   & $P_{\text {max}}$   & $n_{\text {min}}$   & $n_{\text {max}}$   \\ \hline
        ${}_{32}^{72}$Ge  & $P_{\text{ion}}$    & $8.41\cdot 10^{-4}$ & $n_{\text{ion}}$    & $6.23\cdot 10^{-4}$ \\
        ${}_{37}^{85}$Rb  & $8.42\cdot 10^{-4}$ & $1.18\cdot 10^{-3}$ & $6.41\cdot 10^{-4}$  & $7.48\cdot 10^{-4}$ \\
        ${}_{37}^{87}$Rb  & $1.19\cdot 10^{-3}$ & $4.54\cdot 10^{-3}$ & $7.69\cdot 10^{-4}$ & $1.46\cdot 10^{-3}$ \\
        ${}_{50}^{136}$Sn  & $4.55\cdot 10^{-3}$ & $5.49\cdot 10^{-3}$ & $1.71\cdot 10^{-3}$ & $1.87\cdot 10^{-3}$ \\
        ${}_{50}^{140}$Sn  & $5.50\cdot 10^{-3}$ & $8.09\cdot 10^{-3}$ & $1.93\cdot 10^{-3}$ & $2.33\cdot 10^{-3}$ \\
        ${}_{46}^{136}$Pd  & $8.10\cdot 10^{-3}$ & $1.36\cdot 10^{-2}$ & $2.45\cdot 10^{-3}$ & $3.14\cdot 10^{-3}$ \\
        ${}_{58}^{196}$Ce  & $1.37\cdot 10^{-2}$ & $1.43\cdot 10^{-2}$ & $3.63\cdot 10^{-3}$ & $3.71\cdot 10^{-3}$ \\
        \hline\end{tabular*}
\end{center}

\begin{center}
\includegraphics[width=9cm]{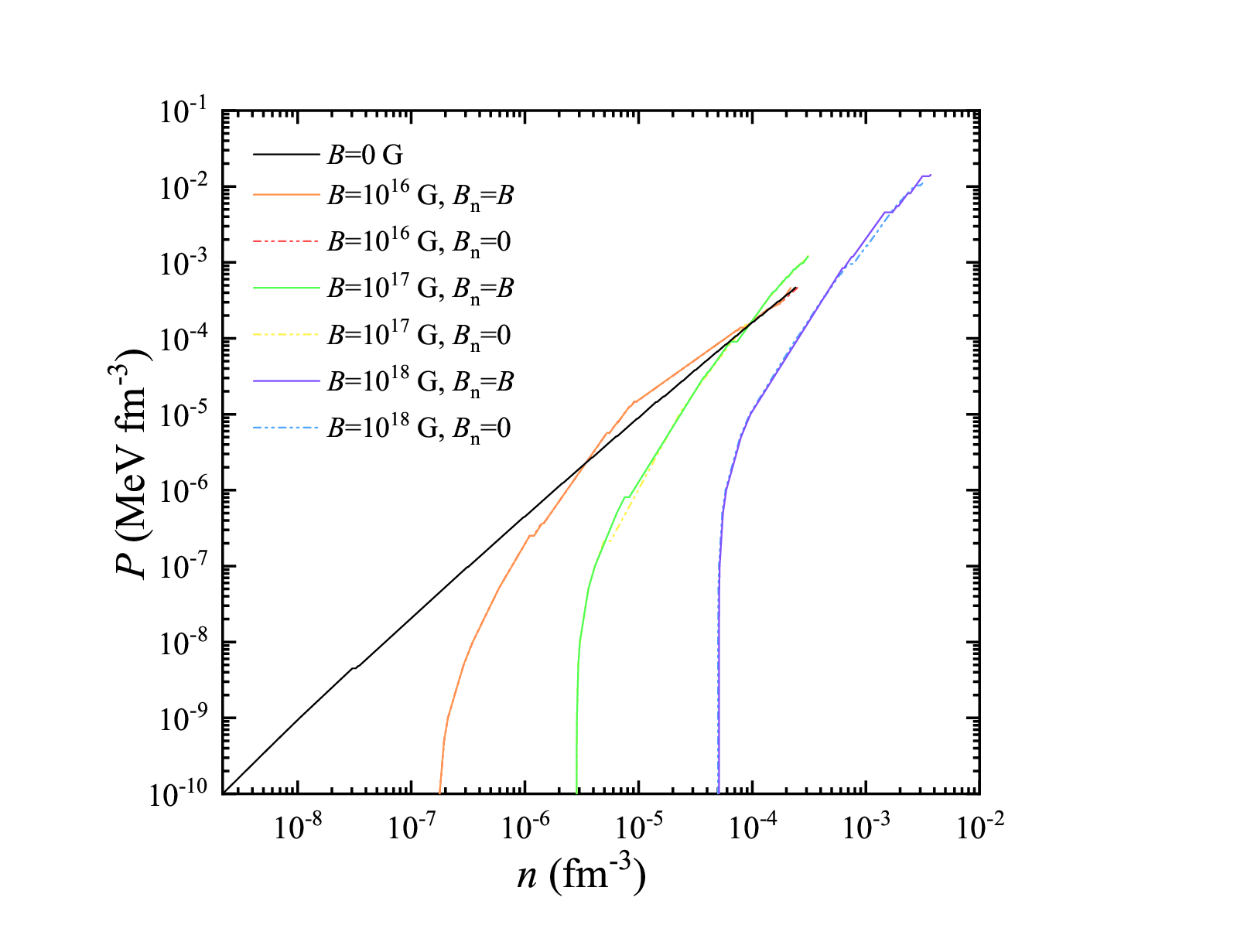}
\figcaption{\label{figg10} Pressure $P$ versus average nucleon
number density $n$ (equation of state) in the outer crust of a cold
nonaccreting neutron star at four magnetic fields $B = 0\ \text G$,
$B = {10^{16}}\ \text G$, $B = {10^{17}}\ \text G$, and $B =
{10^{18}}\ \text G$, where the effect of the magnetic fields on the
nuclear masses is included ($B_{\text n}=B$) and not included
$(B_{\text n}=0)$.}
\end{center}

\begin{center}
\includegraphics[width=9cm]{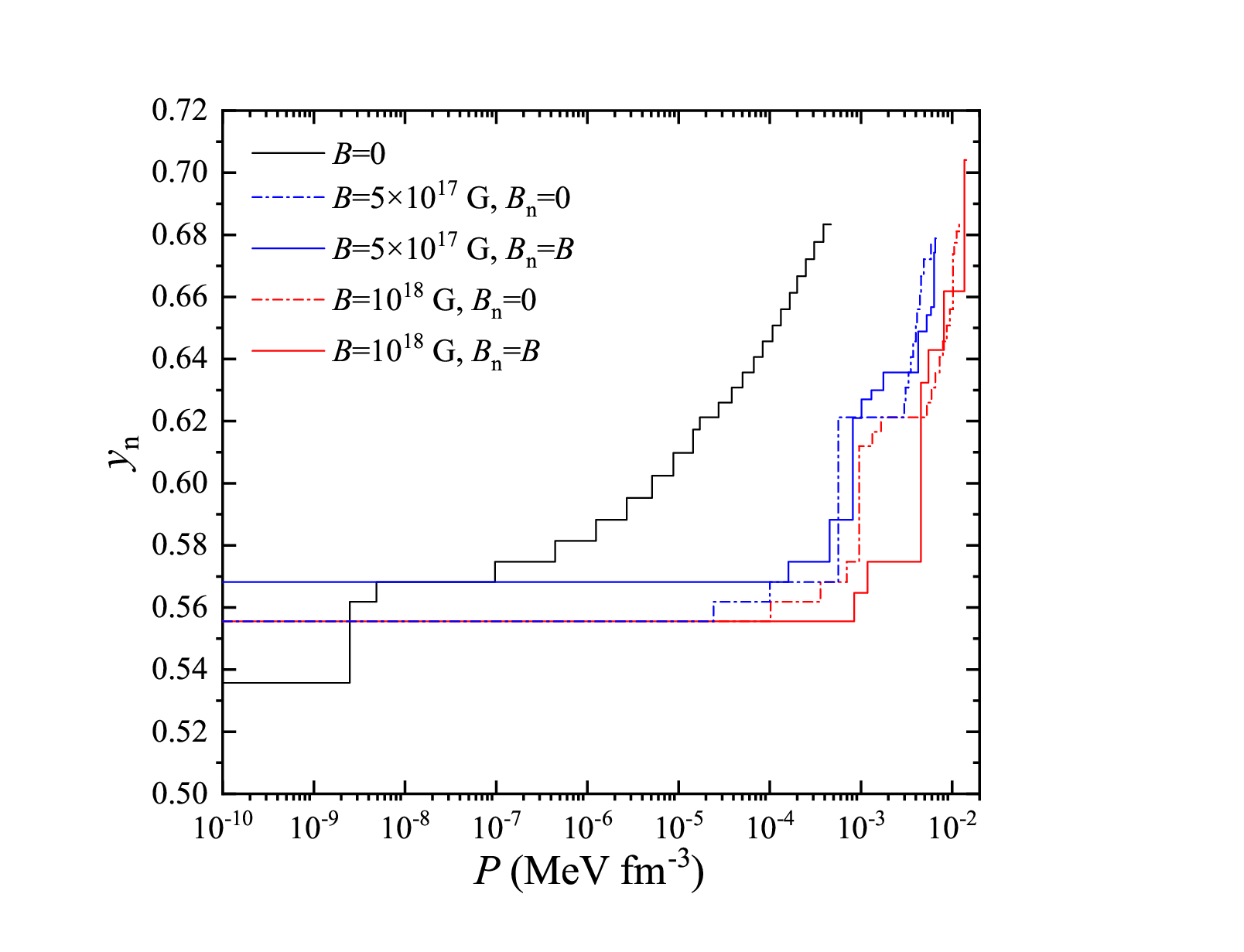}
\figcaption{\label{figg11} Neutron fraction $y_{\text n}$ ($y_{\text
n}$=$N/A$) versus pressure $P$ in the outer crust of a neutron star
for the magnetic field strength of $B=0\ \text G$, $B = {5\times
10^{17}}\ \text G$ and $B = {10^{18}}\ \text G$, where the effect of
the magnetic fields on the nuclear masses is included ($B_{\text
n}=B$) and not included $(B_{\text n}=0)$. }
\end{center}

In Fig. \ref{figg11}, we compare the neutron fraction $y_{\text n}$
in the outer crust for the cases $B=0$ G, $5\times10^{17}$ G, and
$10^{18}$ G. It is shown that $y_{\text n}$ is positively correlated
with the pressure $P$ in all cases. This is consistent with the
physical intuition as the inverse $\beta\text{-}$decay reaction
becomes more intense with increasing depth of the magnetar, and the
nuclei will become more and more neutron-rich. For $B \neq 0$ G, it
is shown that no essential difference can be seen between $B_{\text
n}= 0$ and $B_{\text n}= B$ cases, and $y_{\text n}$ is smaller than
that in the $B=0$ G case at the same pressure (these ultra-strong
MF, i.e., $B>10^{17}$ G, can make the outermost nuclide become
richer in neutron than $^{56}$Fe, and thus may lead to a
neutron-richer region in the outer crust). It indicates that except
in the outermost region for the ultra-strong MF case, the nuclei in
the outer crust of a magnetar are more symmetric than those in an
ordinary neutron star with weak MF at the same pressure.

\section{ Summary}

In summary, we explore the properties of 4110 nuclides from $Z=5$ to
$Z=82$ with the Sky3D code and the composition of the outer crust of
a cold non-accreting magnetar. The effects of the variation on the
nuclear masses due to the MF have been investigated. The MF play an
important role in the binding energies, size and shape of the
nucleus. The strong MF increase the binding energies and nuclear
radii. For the MF $B \lesssim 10^{17}$ G, the nuclear shapes are
nearly unchanged. At higher MF ($B=10^{18}$ G in this work), almost
all the nuclides are deformed while some nuclei with large
deformation can become more 'spherical', moreover, the prolate and
oblate deformations of a nucleus emerge alternately with increasing
neutron numbers in general. For the MF $B \lesssim 10^{17}$ G, there
are some trends in common with the case in the absence of MF, i.e.,
$^{56}$Fe occurs first, the plateaus with $N=50$ and $82$ are
present, and the nuclide at the neutron-drip transition might not
change, while these trends may be altered for $B> 10^{17}$ G. If the
effects of the MF on the nuclear masses are not taken into account
($B_{\text n} = 0$), the MF only influence the electrons but the
outermost nuclide in the outer crust is $^{90}$Zr, not $^{56}$Fe.
The $N=50$ and $N=82$ platforms are still present. Even the neutron
plateau with new magic number $N=126$ can be seen for $B= 10^{18}$
G. When the variation on the nuclear masses due to the MF is
considered, the spin-up and spin-down states split. Thus the energy
levels of a nucleus may undergo the rearrangement, and the situation
may become more complex. In the case of $B_n=B$, the outermost
nuclide is $^{92}$Nb for $B= 5\times10^{17}$ G and $^{72}$Ge for $B=
10^{18}$ G. Moreover, the $N=82$ and $N=126$ plateaus disappear for
$B= 10^{18}$ G. Thus, some concepts, e.g., magic numbers, should be
reexamined in the $B_n=B$ case. In addition, we also study the
neutron-drip transition pressure, the equation of state and neutron
fraction of the outer crust in the presence of the MF. The effects
of MF on $P_{\text{drip}}$ are almost nothing other than for
$B>10^{16}$ G. At the extreme high MF, $B=10^{18}$ G,
$P_{\text{drip}}$ in the $B_{\text n}=B$ case increases about 20\%
in comparison with that for the case of $B_{\text n}=0$. The
equation of state is affected by the MF mainly in the low density
region, where the density is almost unchanged over a wide range of
pressures while the variation on the nuclear masses due to the MF
plays a negligible role. The neutron fractions do not show the
essential difference between the $B_{\text n}=0$ and $B_{\text n}=B$
cases.

The present work shows that the nuclei with the neutron number
$N\geq126$ emerge in both $B_{\text n}=0$ and $B_{\text n}=B$ cases
for $B= 10^{18}$ G, while the $N=126$ nuclei emerge in $B_{\text
n}=B$ case and not in $B_{\text n}=0$ one for $B= 5\times10^{17}$ G.
Thus, considering the effects of the MF on the nuclear masses or not
may modify the conclusion about the composition of the outer crust
under strong MF. It needs to be studied in more detail later.

\end{multicols}

\vspace{-1mm} \centerline{\rule{80mm}{0.1pt}} \vspace{2mm}

\begin{multicols}{2}

\end{multicols}

\vspace{10mm}

\clearpage

\end{CJK*}
\end{document}